\documentclass{article}
\usepackage{amssymb,amsmath,amsthm,proof}
\usepackage{graphicx}
\usepackage{gensymb}

\usepackage{multicol}
\usepackage{multirow}

\newtheorem{thm}{Theorem}[section]

\newtheorem{example}{Example}[section]
\newtheorem*{remark}{Remark}

\usepackage[section]{algorithm}
\usepackage{algpseudocode,algorithmicx}

\newcommand*\Let[2]{\State #1 $\gets$ #2}

\def\Z{{\mathbb Z}}

\title{
A CRT Framework for Montgomery-Type Modular Reduction}
\date{}
\author
{Guangwu Xu\thanks{SCST, Shandong University, China,  e-mail: {\tt gxu4sdq@sdu.edu.cn}(Corresponding author)},
Yanze Yang\thanks{SCST, Shandong University, China,  e-mail: {\tt yyannze99@mail.sdu.edu.cn}.}, Yiran Jia\thanks{SCST, Shandong University, China, e-mail: {\tt  kawyon@mail.sdu.edu.cn}.}.\\
}

\begin{document}


\baselineskip18pt


\maketitle

\begin{abstract}	
Montgomery reduction is one of the fundamental techniques for efficient modular arithmetic. In this paper, we present a new interpretation of Montgomery-type reduction algorithms through the Chinese Remainder Theorem (CRT). We show that the classical Montgomery reduction algorithm arises naturally from the CRT identity, which further reveals a common algebraic invariant underlying a family of Montgomery-type reduction algorithms. This leads to a unified CRT framework for their derivation, analysis, and verification. Within this framework, several recent variants of Montgomery reduction are interpreted in a uniform manner, their correctness proofs become transparent, and their differences are seen to lie only in the representation of the correction term and the evaluation of a common CRT quotient. The framework also provides a convenient tool for analyzing existing reduction algorithms, allowing incorrect parameter ranges to be identified and counterexamples to be constructed naturally.


\end{abstract}

\section{Introduction}
Modular arithmetic is one of the most fundamental computational operations in number theory and plays an indispensable role in both theory and applications. With the rapid development of computing and communication technologies, efficient algorithms for modular arithmetic have become a central topic in computational number theory and cryptography. For example, the basic operations in finite-field arithmetic, RSA, and elliptic curve cryptography all rely heavily on modular multiplication, and improving its efficiency has attracted extensive research. Modular reduction also appears repeatedly in post-quantum cryptography and fully homomorphic encryption, particularly in the implementation of number-theoretic transforms.

For a fixed modulus $n>0$, the most basic modular operation is the computation of the modular product
\[
ab\pmod n
\]
where $0\le a, b<n$. The straightforward approach is based on division with remainder. For any integer $d$ and any positive integer $n$, there exist unique integers $q$ and $r$ such that
\[
d=qn+r, \quad \ 0\le r < n.
\]
The remainder $r$ is precisely
$d \pmod n$. Thus, modular reduction is essentially a division-with-remainder operation, whose computational cost is generally much higher than that of multiplication. 

The situation becomes considerably simpler when the modulus is a power of the radix used by the underlying numeral system, where the quotient and remainder are obtained just by digit shifting or truncation. This simple observation, however, applies only when the divisor is a power of the radix. For a general divisor, the above shortcut no longer works. To overcome this limitation, the Chinese mathematician Yang Hui (1238--1298) proposed several techniques that replace division by multiplication with a carefully chosen constant. The underlying principle is simple: for certain divisors $n$, one finds an integer $c$ such that
$cn=10^s$, or more generally $cn$ is an approximation of $10^s$. The quotient, or an approximation to it, then can be obtained conveniently from 
\[
\frac{d}{n} = \frac{d\times c}{10^s} , \quad \mbox{ or }\quad \frac{d}{n} \approx \frac{d\times c}{10^s}.
\]
In Yang Hui's terminology,  this may be summarized as ``multiply by $c$ instead of dividing by $n$''.


The general strategy of replacing division by multiplication have been developed much further in modern computational number theory.
The celebrated Montgomery reduction transforms modular reduction into a division by a power of the radix, thereby replacing the expensive division by inexpensive shift operations. Another well-known algorithm, Barrett reduction, closely parallels Yang Hui's strategy through the replacement of division by multiplication using a precomputed approximation to its reciprocal.

Among all known techniques for modular multiplication, Montgomery's algorithm is undoubtedly one of the most successful. Introduced by Montgomery in 1985 \cite{Mont}, it is an ingenious design that makes a very astute use of the fact that integer division (and multiplication) by a power of $2$ is trivial.
Given an odd integer $N>0$, one chooses an $R>N$ which is a power of $2$. 
Montgomery first introduced a clever representation of the residue ring 
\[
\Z_n=\Z/n\Z
\]
under which modular multiplication can be carried out remarkably efficiently.

For an integer $a\in [0,n-1]$, its Montgomery representation is given by
\[
\check{a} = aR\pmod n.
\]
Furthermore, for an integer $T\in[0, Rn-1]$, define its Montgomery reduction by
\[
Redc(T)=TR^{-1} \pmod n
\]

The mapping
\[
a\mapsto \check{a}
\]
is clearly a bijection because $\gcd(R,n)=1$. Moreover, addition is preserved:
\[
\check{a}+\check{b}\equiv \check{c}\pmod n \iff a+b\equiv c\pmod n.
\]
However, the multiplication behaves differently. Suppose
$ c=ab\pmod n$, then 
\[
\check{a}\check{b}\equiv abR^2 \equiv \check{c}R\pmod n
\]
Thus one extra factor of $R$ must be removed after each multiplication, which is exactly the purpose of Montgomery reduction.
More specifically, with  exactly the same formulation as in  \cite{Mont}, Montgomery's reduction algorithm  REDC computes $TR^{-1}\pmod N$ for
$0\le T<RN$ and  $N' = R-(N^{-1} \mod R)$:

{\bf function}   REDC($T$) \\
${}\quad\quad (1)\quad\quad m\gets (T \mod R) N' \mod R$\\
${}\quad\quad (2)\quad\quad t\gets \dfrac{T +mN}R$\\
${}\quad\quad (3)\quad\quad${\bf if } $t\ge N$ {\bf then return } $t-N$ {\bf else return } $t \qquad \blacksquare$

The key feature of REDC is that the only division is by $R$, a power of two, which can be implemented by a simple shift operation.

For a technical motivation of Montgomery reduction, the literature often regards it as a generalization of Hensel's odd-division procedure for $2$-adic numbers \cite{BM,SV}. More recently, the extensive use of the number-theoretic transform (NTT) in post-quantum cryptography and fully homomorphic encryption has led to several Montgomery-type modular reduction algorithms tailored to specific applications; see, for example, \cite{AMOT,Huang,Huang2,Plan,Seiler}. These variants retain the Montgomery representation but modify the original reduction procedure in various ways, such as using signed representatives to shift or enlarge the output range, restricting the admissible input range, or reorganizing the arithmetic to suit particular machine architectures.
The correctness of these variants is typically established by a direct, step-by-step verification. Although such proofs are valid and often technically delicate, they tend to be algorithm-specific and provide limited insight into why the particular modifications work or how different variants are related. In particular, the underlying algebraic mechanism that explains both the construction of the algorithms and the correctness of their output ranges is usually left implicit.

These developments naturally raise a fundamental question. Is Montgomery reduction merely an ingenious computational trick, together with a collection of independently designed variants, or is there a common mathematical principle from which these algorithms can be systematically derived? Answering this question is not only of theoretical interest, but also provides a conceptual foundation for the design and verification of new Montgomery-type algorithms.

This paper answers this question by showing that Montgomery reduction admits a natural formulation in terms of the Chinese Remainder Theorem (CRT), so that the number theoretic nature and computational characteristic can be seen more transparently. More specifically, 
our contributions are three-fold.
\begin{enumerate}
	\item  We derive Montgomery reduction directly from a CRT reconstruction identity, from which the seemingly ad hoc quantity $m= (T \mod R) N' \mod R$ arises naturally as the coefficient of the $N$-component.
	This is a straightforward process that has a clear algebraic motivation.
	\item  We establish a general CRT framework for Montgomery-type reductions, which unifies many existing variants. Instead of proving each algorithm correct by ad hoc manipulations of congruences and inequalities, the CRT framework
	derives Montgomery-type algorithms by construction from a common CRT reconstruction principle.
	\item  The framework provides a systematic validation tool for newly proposed Montgomery-type algorithms. Correctness becomes a consequence of the CRT construction, while incorrect designs can be detected immediately.
\end{enumerate}

This paper describes a new CRT interpretation and a unifying framework for Montgomery-type reduction algorithms. Although it suggests a set of general principles for designing such variants, it does not present a new algorithm.

This paper is a substantially revised version of our earlier preprint \cite{YJX}. The present version focuses exclusively on modular reduction and the CRT framework, while omitting the applications to the number theoretic transform. 
It is remarked that the first author has presented this viewpoint in several invited and public seminar talks since 2021\footnote{For example, one of which was given at the IIE of Chinese Academy of Sciences on November 25, 2021.}.

In the next section, we begin with Qin's identity, from which a CRT reconstruction formula is obtained immediately. We then show that a simple manipulation of this formula leads directly to Montgomery reduction.
This observation naturally leads to a general CRT framework for Montgomery-type modular reduction. Section 3 applies the framework to several recent variants of Montgomery reduction, where their correctness follows uniformly from the CRT formulation. Section 4 concludes the paper.

\section{Qin's Identity, CRT and Montgomery Algorithm}\label{sec2}
As mentioned before, Montgomery reduction is traditionally motivated as a generalization of Hensel's odd division for $2$-adic numbers. In this section we present a different viewpoint. We observe that the algebraic expression underlying Montgomery reduction is naturally obtained from the Chinese Remainder Theorem (CRT). More precisely, a classical identity already suggested in Qin Jiushao's treatment of the CRT immediately yields the Montgomery reduction formula. This viewpoint also explains several modern variants of Montgomery reduction in a unified and transparent way.

In this section, we start with some historical notes related to the Chinese remainder theorem.
The CRT is a well known method for solving a system of modular equation.
It is also known as  Sun Tzu Theorem as it was described in a very ancient Chinese book ``Sun Tzu Suan Jing''.
In  his  ``Mathematical Treatise in Nine Sections'' of 1247 \cite{qin1247},  Qin
described the Chinese remainder theorem with great detail and generality.

In \cite{qin1247}, Qin introduced the notion of ``precise use'' (``ch\^eng-yung'')\footnote{In \cite{Libb} (page 331), `precise use' is termed as `ch\^eng-yung' to keep Chinese pronunciation.}, whose essential algebraic content, in the case of two moduli, is the following identity. We remark that
although its proof is straightforward, the fact that Qin paid special attention to this expression
in his derivation of CRT is very interesting.

\begin{thm} (Qin's Identity)\label{qin:iden} The following identity is an algebraic formulation of Qin's ``precise use'' for two moduli: let
	$N, R>1$ be two coprime integers. Denote $N^{-1} = N^{-1}\pmod R, \ R^{-1}=R^{-1}\pmod N$,
	then
	\begin{equation}\label{eq:qin0}
		R^{-1}R+N^{-1}N = 1+NR.
	\end{equation}
\end{thm}
\begin{proof}
	Since
	\[
	N^{-1}N \equiv 1 \pmod R,
	\]	
	there exists a positive integer $\ell$ satisfying
	\[
	N^{-1}N = 1+\ell R.
	\]
	Note that  $1\le N^{-1}<R$, we see that $\ell <N$. Furthermore, $\ell R\equiv -1 \pmod N$, so $N-\ell=R^{-1}$. Therefore,
	\[
	R^{-1}R +N^{-1}N=(N-\ell)R+N^{-1}N=1+NR.
	\]
\end{proof}

Multiplying both sides of (\ref{eq:qin0}) by an arbitrary integer $T$, we obtain
\begin{equation}\label{eq:qin1}
	TR^{-1}R+TN^{-1}N = T+TNR.
\end{equation}

Before turning to Montgomery reduction, we briefly recall that (\ref{eq:qin1}) immediately yields the classical CRT reconstruction formula.
For the system of congruence 
\[
\left\{ \begin{array}{l} T \equiv r_1 \pmod N\\
	T\equiv r_2\pmod R \end{array}\right.
\]
We have 
\[
TR^{-1}R+TN^{-1}N\equiv r_1R^{-1}R+r_2N^{-1}N \pmod {NR}.
\]
By
(\ref{eq:qin1}) we obtain the CRT solution formula:
\[
T\equiv r_1R^{-1}R+r_2N^{-1}N \pmod {NR}.
\]
For the general case of multiple moduli, one can derive the CRT 
in a similar manner, see \cite{dlx}. 

At this point the connection with Montgomery reduction becomes immediate.
Writing $N'=R-N^{-1}$,  the identity (\ref{eq:qin1})  becomes
\begin{equation}\label{eq:qin2}
	TR^{-1}R=T+TN'N
\end{equation}
This means that $T+TN'N$ is divisible by $R$ and 
\[
TR^{-1}=\frac{T+TN'N}R
\]
Recall that $m = TN' \pmod R = (T \pmod R)N' \pmod R$ in the  Montgomery reduction,
so
\begin{equation}\label{eq:qin3}
	TR^{-1}\equiv \frac{T+mN}R \pmod N.
\end{equation}
Here $m$ replaces the exact product $TN'$ by its residue modulo $R$, which does not affect the quotient modulo $N$ because the difference is a multiple of $RN$.

Since 
\[
0\le m<R, \quad 0\le T<RN,
\]
we have
\[
\frac{T+mN}R=\frac{T}R+\frac{mN}R<2N,
\]
i.e., (\ref{eq:qin3}) gives the celebrated Montgomery reduction algorithm.

This derivation differs fundamentally from the traditional $2$-adic motivation. Rather than introducing Montgomery reduction through Hensel lifting, it arises directly from one term of the CRT identities (\ref{eq:qin1}) and (\ref{eq:qin2}). In this sense, the Montgomery formula is not merely analogous to the CRT---it is literally obtained by rearranging the CRT identity.

Some general principles are suggested from the process. To this end, we first fix a notation {\bf mods} for signed remainder in this paper:  by $x ( \mbox{ mods } n)$, we mean the least absolute remainder of an integer $x$ dividing by a positive integer $n$ (which is an integer in the interval $[-\frac{n}2, \frac{n}2)$). 

The above derivation suggests that the essential ingredients of Montgomery reduction are independent of the specific choice of $m$. This leads naturally to the following general framework encompassing Montgomery reduction and several of its variants.
The task for Montgomery algorithm (and its variants) is to efficiently compute
$r= TR^{-1} \pmod N$, or $r=  TR^{-1} (\mbox{ mods } N)$, or
$r\equiv TR^{-1} \pmod N$ for integer $r$ in certain range (e.g., $a\le r<b$). By (\ref{eq:qin1}), that means that
\begin{equation}\label{eq:qin4}
	r  \equiv \frac{T-m N}R \pmod N , \quad  m = TN^{-1} \pmod R \, \mbox{ or }
	m=  TN^{-1} (\mbox{ mods } R)
\end{equation}
Some of the key points are
\begin{itemize}
	\item $T-m N$ is divisible by $R$.
	\item The integer $m$ should be chosen so that $\dfrac{T-mN}R$ remains small.
	Given that $|T| <RN$, this is true for $m= TN^{-1} \pmod R$, or $m=  TN^{-1} (\mbox{ mods } R)$.
	\item The returned value is the quotient $\dfrac{T-mN}R$, whose computation can often be accelerated by exploiting the structure of $R$.
\end{itemize}
Some Montgomery-type algorithms instead compute
\[
-TR^{-1} \pmod N \quad \mbox{or} \quad - TR^{-1} (\mbox{ mods } N).
\]
In this case one simply returns
\[
\frac{m N - T}R,
\]
which follows from the same framework.
\begin{remark}
	The derivation above shows that Montgomery reduction is not merely compatible with the Chinese Remainder Theorem. Rather, it is obtained by an algebraic rearrangement of the CRT reconstruction identity. From this perspective, the CRT reconstruction formula and Montgomery reduction are two different realizations of the same underlying algebraic invariant. This invariant provides a unified framework for understanding Montgomery reduction and its variants.
\end{remark}

\section{A Unified CRT Analysis of Montgomery-Type Reduction Algorithms}
The number theoretic transform (NTT) has become a fundamental computational tool in lattice-based cryptography, especially for schemes based on structured polynomial lattices. 
Consequently, efficient modular multiplication for small (typically word-size) moduli has attracted considerable attention, leading to several variants of Montgomery reduction \cite{AMOT, Huang,Huang2,LyuSei,Plan,Seiler}.

The previous section shows that Montgomery reduction is a direct consequence of the CRT identity. This observation extends well beyond the original Montgomery algorithm. In the remainder of this paper we show that several recently proposed Montgomery-type reduction algorithms all arise from the same identity. Consequently, their correctness proofs become uniform, while implementation details become considerably easier to analyze. Throughout this section, we use notation consistent with the classical Montgomery algorithm.
We shall analyze signed Montgomery reduction, 	Plantard reduction and signed Plantard  reduction in this section.

Table~\ref{tab:CRTFramework} summarizes the common structure of the algorithms analyzed in this section. Although their implementations differ, each algorithm computes the same quotient arising from the CRT identity; they differ only in the representation of the correction term $m$ and in the evaluation of the quotient.

\begin{table}[ht]
	\centering
	\caption{CRT framework for Montgomery-type reduction algorithms.}
	\label{tab:CRTFramework}
	\renewcommand{\arraystretch}{1.2}
	\begin{tabular}{|l|c|l|l|}
		\hline
		\textbf{Algorithm} &
		\textbf{Universal Quotient} &
		\textbf{Choice of $m$} &
		\textbf{Evaluation technique} \\
		\hline
		
		Montgomery &
		$\displaystyle \frac{T+mN}{R}$ &
		$m\equiv TN' \pmod R$ &
		direct \\
		\hline
		
		Signed Montgomery &
		$\displaystyle \frac{T-mN}{R}$ &
		$m\equiv TN^{-1}\;(\mathrm{mods}\;R)$ &
		$T=a_1R+a_0$ \\
		\hline
		
		Plantard &
		$\displaystyle \frac{mN-T}{R}$ &
		$m\equiv TN^{-1}\pmod R$ &
		high-half extraction \\
		\hline
		
		Signed Plantard &
		$\displaystyle \frac{mN-T}{R}$ &
		$m\equiv TN^{-1}\;(\mathrm{mods}\;R)$  &
		rounding \\
		\hline
	\end{tabular}
\end{table}

The remainder of this section demonstrates that every Montgomery-type reduction algorithm computes exactly one of the two universal CRT quotients
\[
\frac{T-mN}{R}\quad \mbox{or} \quad \frac{mN-T}{R}
\]
with different algorithms differing only in the choice of the correction term m and in the implementation of the quotient evaluation.

\subsection{CRT Interpretation of Signed Montgomery Algorithm}
Recently, Seiler proposed a variation Montgomery reduction and used it to perform NTT for NTRU
\cite{Seiler,LyuSei}. The modulus $R=2^{\ell}$
is restricted in the size of a word and
negative residue is allowed. The odd modulus $N$ satisfies $2N< R$.
The input $T$ with $|T|<\dfrac{NR}2$ is written as $T=a_1R+a_0$ with $0\le a_0<R$.
The description of the Seiler algorithm is next.

\begin{algorithm}[htb]
	\caption{Signed-Montgomery Reduction
		\label{alg-seiler}}
	\begin{algorithmic}[1]
		\Require{Integer $ -\dfrac{NR}2<T=a_1 R+a_0< \dfrac{NR}2, \ 0\le a_0<R$}
		\Ensure{ $ r\equiv TR^{-1} \pmod N$, $-N<r<N$.}
		\Function{SigRedc}{$T$}
		\Let{$m_0$}{$a_0N^{-1} ( \mbox{ mods } R)$ }
		\Let{$t$}{$\left\lfloor\dfrac{m_0N}{R}\right\rfloor$ }
		\Let{$r$}{$a_1-t$}
		\algstore{Sign}
	\end{algorithmic}
\end{algorithm}

\begin{algorithm}[htb]
	\begin{algorithmic}[1]
		\algrestore{Sign}
		\State \Return{$r$}
		\EndFunction
	\end{algorithmic}
\end{algorithm}

Next we present a straightforward derivation of Algorithm \ref{alg-seiler}, together with a proof of its  correctness.

As discussed at the end of last section, the return value $r$ satisfies
\[
r\equiv TR^{-1} \pmod N \equiv \frac{T-mN}R \pmod N 
\]
with $m \equiv TN^{-1} ( \mbox{ mods } R)$. Since $T$ has a specific representation
of $T=a_1R+a_0$ with $0\le a_0<R$, we get 
\[
m  \equiv (a_1R+a_0)N^{-1} ( \mbox{ mods } R) \equiv a_0N^{-1} ( \mbox{ mods } R)=m_0.
\]
This implies that $m=m_0$ as they are all in $[-\frac{R}2, \frac{R}2)$.
Now
\[
\frac{T-mN}R \pmod N  \equiv \frac{a_1R+a_0-mN}R \pmod N =a_1- \frac{m_0N-a_0}R=r.
\]
This gives the desired Signed-Montgomery Reduction algorithm simply because
\begin{enumerate}
	\item $\left| a_1-\dfrac{mN-a_0}R \right|=\left| \dfrac{a_1R+a_0-mN}R \right|=\left| \dfrac{T-mN}R \right|
	<\frac{\dfrac{NR}2+\dfrac{R}2 N}R=N$, since $|m|\le \dfrac{R}2$.
	\item $\left\lfloor\dfrac{mN}{R}\right\rfloor = \dfrac{mN-a_0}R$, since $\dfrac{mN-a_0}R$ is an integer and $0\le \dfrac{a_0}R<1$.
\end{enumerate}

Algorithm 3.1 is simply the evaluation of the universal quotient
\[
\frac{T-mN}R \pmod N
\]
under the decomposition
\[
T=a_1 R+a_0.
\]

\subsection{CRT Interpretation of Plantard Reduction Algorithm and Its Extension}
In 2021,  Plantard designed a new modular reduction algorithm \cite{Plan}.
The method utilizes the special property of doing modular operation with in a word size. It
exhibits better performance for several cryptographic schemes. One of the nice features of the algorithm is to turn a multiplication of a $2n$-bit integer with an $n$-bit integer to a multiplication of two   $n$-bit integers.
To be more precise, let $\phi = \dfrac{1+\sqrt{5}}{2}$, $R=2^{2n}$. 
If the odd modulus $N$ satisfies $N<\dfrac{2^n}{\phi}$, then the following algorithm works correctly.

\begin{algorithm}[htb]
	\caption{Plantard Reduction Algorithm
		\label{alg-Plan}}
	\begin{algorithmic}[1]
		\Require{$N<\frac{2^n}{\phi},0 \leq T \leq N^2$, }
		\Ensure{$r = -TR^{-1}\pmod N$.}
		\Function{PRedc}{$T$}
		\Let{$r$}{$\left\lfloor\dfrac{\left(\left\lfloor\dfrac{ T N^{-1} \pmod R }{2^n}\right\rfloor +1 \right) N}{2^n}\right\rfloor$ }
		%
		\If {$(r = N)$}
		\State \Return{$0$}
		\EndIf
		\State \Return{$r$}
		\EndFunction
	\end{algorithmic}
\end{algorithm}

To use the CRT model, we argue as follows.

Writing $m=TN^{-1}\pmod R$.  From (\ref{eq:qin1}), we get
\[
-TR^{-1}R = TN^{-1}N-T-TNR=\big(TN^{-1} \pmod R\big)N-T+kNR
\]
for some integer $k$. This means that $\big(TN^{-1} \pmod R\big)N-T$ is divisible by $R$ and
\[
-TR^{-1}\equiv \frac{mN-T}R\pmod N
\]
Therefore the only remaining task is to prove that the implementation computes exactly the quotient $\dfrac{mN-T}R$, i.e.,
\begin{equation}\label{eq:Pl}
	\frac{mN-T}R=\left\lfloor\frac{\left(\left\lfloor\frac{ m }{2^n}\right\rfloor+1\right) N}{2^n}\right\rfloor.
\end{equation}
Note that $R=2^{2n}$ and $0\le T<N^2<\frac{R}{\phi^2}$, so (\ref{eq:Pl}) is equivalent to 
\[
\frac{mN-T}{2^n}\le \bigg(\left\lfloor\frac{m }{2^n}\right\rfloor+1\bigg) N<\frac{mN+R-T}{2^n}.
\]
The first inequality is obvious. By the fact that $\dfrac{1}{\phi}=1-\dfrac{1}{\phi^2}$, we get   the second inequality:
\begin{eqnarray*}
	\frac{mN+R-T}{2^n} &\ge & \frac{mN+R-\frac{R}{\phi^2}}{2^n}=\frac{m }{2^n}N+2^n(1-\frac{1}{\phi^2})\\
	&=&\frac{m }{2^n}N+\frac{2^n}{\phi}>\bigg(\left\lfloor\frac{m }{2^n}\right\rfloor+1\bigg) N.
\end{eqnarray*}
Finally, since $0\le m<R, \ 0\le T <R$,  it is clear that $-1<\frac{mN-T}R<N$. Namely,
\[
0\le \frac{\big(TN^{-1} \pmod R\big)N-T}R<N.
\]

{\bf Remark}. It is interesting to note that we have actually shown that
\[
\left\lfloor\dfrac{\left(\left\lfloor\dfrac{ T N^{-1} \pmod R }{2^n}\right\rfloor+1\right)N}{2^n}\right\rfloor<N ,
\]
so the {\bf if} clause
in lines 3 and 4 of algorithm \ref{alg-Plan} can be removed. We remark that this has been pointed out in \cite{AMOT} through some calculation.

A further discussion on Plantard Reduction Algorithm is given in  \cite{AMOT}  where a signed version is proposed. The 
algorithm considers $R=2^{2n}$ and odd modulus $N$ such that $N<2^{n-1}$ and sets $\widetilde{N}^{-1} = N^{-1}  (\mbox{ mods } R),  \   \widetilde{R}^{-1} = R^{-1}  (\mbox{ mods } N)$. The goal is to
compute  $-T\widetilde{R}^{-1} (\mbox{ mods } N)$
for an input $T$ with $|T|<2^{2n-2}$.

\begin{algorithm}[htb]
	\caption{Signed Plantard Multiplication
		\label{alg-SigPlan}}
	\begin{algorithmic}[1]
		\Require{Integer $T$ with $ |T|\leq 2^{2n-2}$ }
		\Ensure{$r = -T\widetilde{R}^{-1}(\mbox{ mods } N)$.}
		\Function{SigPRedc1}{$T$}
		\Let{$r$}{$\left\lfloor\dfrac{\left\lfloor\dfrac{ T \widetilde{N}^{-1} (\mbox{ mods } R) }{2^n}\right\rceil N}{2^n}\right\rceil$ }
		\State \Return{$r$}
		\EndFunction
	\end{algorithmic}
\end{algorithm}

Applying the CRT framework again, we rewrite (\ref{eq:qin1})  as $T\widetilde{R}^{-1} R+T\widetilde{N}^{-1} N=T+kNR$ for some integer $k$. Therefore, $\big(T\widetilde{N}^{-1} (\mbox{ mods } R) \big)N-T$ is divisible by $R$ (with a smaller quotient) and 
\[
-T\widetilde{R}^{-1}(\mbox{ mods } N)\equiv \frac{\big(T\widetilde{N}^{-1} (\mbox{ mods } R) \big)N-T}{R}.
\]
Write $m=T\widetilde{N}^{-1} (\mbox{ mods } R)$\footnote{We use the notation $\widetilde{N}^{-1}$ from \cite{AMOT}. But in fact this number is $m\equiv TN^{-1}\;(\mathrm{mods}\;R)$. },  it therefore suffices to prove that 
\[
\frac{mN-T}{R}=\left\lfloor\frac{\left\lfloor\frac{ m }{2^n}\right\rceil N}{2^n}\right\rceil
\]
Since $=\left\lfloor\dfrac{\left\lfloor\dfrac{ m }{2^n}\right\rceil N}{2^n}\right\rceil=\left\lfloor\dfrac{\left\lfloor\dfrac{ m }{2^n}+\frac{1}2\right\rfloor N}{2^n}+\dfrac{1}2\right\rfloor$, this is equivalent to checking
\[
\frac{mN-T}{2^n}\le \left\lfloor\frac{ m }{2^n}+\frac{1}2\right\rfloor N+2^{n-1}<\frac{mN+R-T}{2^n}
\]
To prove the first inequality, we use the facts of $N<2^{n-1}$ and $|T|<2^{2n-2}$:
\[
\frac{mN-T}{2^n}=\big(\frac{m}{2^n}-\frac{1}2\big)N+\frac{N}2-\frac{T}{2^n}\le \left\lfloor\frac{ m }{2^n}+\frac{1}2\right\rfloor N
+\frac{N}2+\frac{|T|}{2^n}< \left\lfloor\frac{ m }{2^n}+\frac{1}2\right\rfloor+2^{n-1}.
\]
To prove the second inequality, we note that $R-T-2^{n-1}N>2^{2n}-2^{2n-2}-2^{2n-2}=2^{2n-1}$, so  
\[
\left\lfloor\frac{ m }{2^n}+\frac{1}2\right\rfloor N+2^{n-1}\le
\big(\frac{ m }{2^n}+\frac{1}2\big) N+2^{n-1}<\frac{mN+R-T}{2^n}.
\]

\subsection{Discussion on Another Signed Version of Plantard Algorithm}
Observing that a modulus in many applications is usually much smaller than a word, Huang et al introduced another signed version of  Plantard algorithm 
to processes signed integers \cite{Huang}.  In the algorithm, $R$ is set to be
$2^{2n}$, while
the odd modulus $N$ satisfies $N<2^{n- \alpha -1}$, where $\alpha \geq 0$ is an integer parameter. Write $\widetilde{N}^{-1} = N^{-1}  (\mbox{ mods } R),  \   \widetilde{R}^{-1} = R^{-1}  (\mbox{ mods } N)$, the goal is to
compute  $-T\widetilde{R}^{-1} (\mbox{ mods } N)$
for an input $T$ with $|T|<2^{2\alpha}N^2$.

\begin{algorithm}[htb]
	\caption{Signed Plantard Reduction Algorithm-(2)
		\label{alg-SigPlan2}}
	\begin{algorithmic}[1]
		\Require{Integer $T$ with $|T|\le 2^{2\alpha}N^2$, integer $\alpha\ge 1$}
		\Ensure{$r \equiv -T\widetilde{R}^{-1}(\mbox{ mods } N),-\dfrac{N}{2} < r <\dfrac{N}{2}$.}
		\Function{SigPRedc2}{$T,\alpha$}
		\Let{$r$}{$\left\lfloor\dfrac{\left(\left\lfloor\dfrac{ T \widetilde{N}^{-1} (\mbox{ mods } R) }{2^n}\right\rfloor +2^\alpha\right)N}{2^n}\right\rfloor$ }
		\State \Return{$r$}
		\EndFunction
	\end{algorithmic}
\end{algorithm}

The CRT framework also serves as a convenient tool for verification. As an illustration, we analyze the Signed Plantard Reduction Algorithm-(2) (algorithm \ref{alg-SigPlan2}).
Again, from (\ref{eq:qin1})  we get  $T\widetilde{R}^{-1} R+T\widetilde{N}^{-1} N=T+kNR$ for some integer $k$. and hence $\big(T\widetilde{N}^{-1} (\mbox{ mods } R) \big)N-T$ is divisible by $R$ (with a smaller quotient). Therefore 
\[
-T\widetilde{R}^{-1}(\mbox{ mods } N)\equiv \frac{\big(T\widetilde{N}^{-1} (\mbox{ mods } R) \big)N-T}{R}.
\]
Write $m=T\widetilde{N}^{-1} (\mbox{ mods } R)$,  we need to verify that
\[
\frac{mN-T}{R}=\left\lfloor\frac{\left(\left\lfloor\frac{ m }{2^n}\right\rfloor +2^\alpha\right)N}{2^n}\right\rfloor.
\]
Namely, the following inequalities need to be proved:
\[
\frac{mN-T}{2^n}\le \bigg(\left\lfloor\frac{m}{2^n}\right\rfloor +2^\alpha \bigg)N<\frac{mN+R-T}{2^n}.
\]
The second inequality is always true by the facts that $R>2^{n+\alpha+1}N$ and $|T|\le 2^{n+\alpha-1}N$:
\[
\frac{mN+R-T}{2^n}> \left\lfloor\frac{m}{2^n}N\right\rfloor+\frac{2^{n+\alpha+1}N - 2^{n+\alpha-1}N }{2^n}> \left\lfloor\frac{m}{2^n}\right\rfloor N+2^\alpha N.
\]
The first one holds for $\alpha \ge 1$ since
\[
\frac{mN-T}{2^n}< \bigg(\left\lfloor\frac{m}{2^n}\right\rfloor +1 \bigg)N+\frac{|T|}{2^n}< \bigg(\left\lfloor\frac{m}{2^n}\right\rfloor +1 \bigg)N+
2^{\alpha-1}N\le \bigg(\left\lfloor\frac{m}{2^n}\right\rfloor +2^{\alpha} \bigg)N
\]

We would like to point it out that the above does not hold for the case of $\alpha = 0$. We construct a counter example to show that
\[
\frac{mN-T}{2^n}\le  \bigg(\left\lfloor\frac{m}{2^n}\right\rfloor +1 \bigg)N
\] 
is not true in general:
\begin{example} Let $n = 6, N=31, \alpha=0$. So $R=2^{2n}=4096$ and $N$ satisfies $N<2^{n-\alpha-1}=32$.
	
	It is calculated that $\widetilde{N}^{-1} = N^{-1}  (\mbox{ mods } R)= -1057$. Now we take inputs $T=-95$, then $m=-1985=(-95)(-1057)  (\mbox{ mods } R)$ and
	\[
	\frac{mN-T}{2^n}= -15\times 64 = -960.
	\]
	But 
	\[
	\bigg(\left\lfloor\frac{m}{2^n}\right\rfloor +1 \bigg)N = -961.
	\]
	
	In this example, the return value of the algorithm is
	\[
	\left\lfloor\frac{\left(\left\lfloor\frac{ T \widetilde{N}^{-1} (\mbox{ mods } R) }{2^n}\right\rfloor +2^0\right)N}{2^n}\right\rfloor=-16,
	\]
	which is not congruent to $-T\widetilde{R}^{-1}$ modulo $N$, and not even in the required interval
	$[-15,15]$. So algorithm \ref{alg-SigPlan2} is incorrect.
\end{example}
The failure occurs precisely because the key inequality required by the CRT quotient no longer holds when $\alpha=0$. Thus the CRT framework pinpoints the exact source of the error rather than merely producing a counterexample.

\begin{remark}
	This example illustrates another advantage of the CRT framework. Once every reduction algorithm is viewed as an implementation of the quotient
	\[
	\frac{mN-T}{R},
	\]
	correctness reduces to verifying a small number of inequalities. Consequently, both the validity for $\alpha>0$ and the failure for $\alpha=0$ become transparent.
	
	This counter example was reported in the early version of our paper
	\cite{YJX}. The original statement of algorithm \ref{alg-SigPlan2}
	in \cite{Huang} included the case of $\alpha=0$ and it was removed after learning \cite{YJX}.
	This is also reflected in their more recent paper\cite{Huang2}. 
\end{remark}

\section{Conclusion}
In this paper we have shown that Montgomery reduction is naturally derived from Qin's identity underlying the Chinese remainder theorem. This viewpoint leads to a unified framework for Montgomery reduction and several of its modern variants. Within this framework, the algorithms arise from the same algebraic identity, while their differences are reflected only in the choice and computation of the correction term. Consequently, their correctness proofs become uniform and transparent. The framework also provides a convenient tool for analyzing new reduction algorithms, illustrated by simplified correctness proofs, the removal of unnecessary implementation steps, and the identification of an incorrect parameter range in a published algorithm. We believe that the CRT viewpoint offers a natural conceptual foundation for the further development of Montgomery-type reduction algorithms.

\section*{Acknowledgement}
This work is partially supported by the National Natural Science Foundation of China (No. 12271306).

During the preparation of this work, the authors used ChatGPT (OpenAI) to improve the readability and language of the manuscript. After using this tool, the authors carefully reviewed and edited the content and take full responsibility for the publication.

\end{document}